\documentclass[prb, aps, superscriptaddress, twocolumn]{revtex4-2}

\usepackage{graphicx}
\usepackage{dcolumn}
\usepackage{bm}
\usepackage{xcolor}
\usepackage{soul}
\usepackage{float} 
\usepackage{hyperref} 
\hypersetup{
	colorlinks=true, 
	citecolor = blue,
}

\usepackage[version=3]{mhchem}

\renewcommand{\cite}[1]{\textsuperscript{\citenum{#1}}}

\begin{document}
	
	
	\title{Achieving superconductivity in infinite-layer nickelate thin films by aluminum sputtering deposition}
	
	
	\author{Dongxin Zhang}
	\affiliation{Laboratoire Albert Fert, CNRS, Thales, Université Paris-Saclay, 91767 Palaiseau, France}
	
	\author{Aravind Raji}
	\affiliation{Laboratoire de Physique des Solides, CNRS, Université Paris-Saclay, Orsay 91405, France\\}
	\affiliation{Synchrotron SOLEIL, L’Orme des Merisiers, BP 48 St Aubin, Gif sur Yvette, 91192, France}
	
	\author{Luis M. Vicente-Arche}
	\affiliation{Laboratoire Albert Fert, CNRS, Thales, Université Paris-Saclay, 91767 Palaiseau, France}
	
	\author{Alexandre Gloter}
	\affiliation{Laboratoire de Physique des Solides, CNRS, Université Paris-Saclay, Orsay 91405, France\\}
	
	\author{Manuel Bibes}
	\affiliation{Laboratoire Albert Fert, CNRS, Thales, Université Paris-Saclay, 91767 Palaiseau, France}
	
	\author{Lucía Iglesias}%
	\email{lucia.iglesias@cnrs-thales.fr}
	\affiliation{Laboratoire Albert Fert, CNRS, Thales, Université Paris-Saclay, 91767 Palaiseau, France}

	\date{\today}
	
	\begin{abstract}
The recent discovery of superconductivity in infinite-layer (IL, ABO$_2$) nickelates has opened a new avenue to deepen the understanding of high-temperature superconductivity. However, progress in this field is slowed by significant challenges in material synthesis and the scarcity of research groups capable of producing high quality superconducting samples. IL nickelates are obtained from a reduction of the perovskite ABO$_3$ phase, typically achieved by annealing using CaH$_2$ as a reducing agent. Here, we present a new method to synthesize superconducting infinite-layer nickelate Pr\textsubscript{0.8}Sr\textsubscript{0.2}NiO\textsubscript{2} thin films using an aluminum overlayer deposited by sputtering as a reducing agent. We systematically optimized the aluminum deposition parameters and obtained superconducting samples reduced either \textit{in situ} or \textit{ex situ} (after air exposure of the precursor ABO$_3$ films). A comparison of their crystalline quality and transport properties shows that \textit{in situ} Al reduction enhances the quality of the superconducting Pr\textsubscript{0.8}Sr\textsubscript{0.2}NiO\textsubscript{2} thin films, achieving a maximum superconducting transition temperature $T_{c}^{onset}$ of 17 K, in agreement with the optimum value reported for this compound. This simple synthesis route, much more accessible than existing methods, offers better control and reproducibility over the topotactic transformation, opening new opportunities to gain insights into the physics of superconductivity in nickelates.

\begin{description}
	\item[Keywords]
	Superconductivity, infinite-layer nickelates, sputtering, \textit{in situ}  reduction
\end{description}
\end{abstract}

	\maketitle

\section{Introduction}
The discovery of high temperature superconductivity in cuprates in 1986\cite{Onnes(1911)} spurred the quest for analogue systems that could shed light on the mechanism underlying unconventional superconductivity. Researchers then identified nickelates as a potential cuprates analogs mainly due to the isoelectronic properties of $Ni^{1+}$ ion and $Cu^{2+}$, both with $3d^{9}$ electron count.\cite{Anisimov(1999),Hansmann(2009)} After three decades of intense research, it was in 2019 when superconductivity was found in infinite-layer (IL) hole-doped Nd\textsubscript{0.8}Sr\textsubscript{0.2}NiO\textsubscript{2} nickelate thin films\cite{Li(2019)} grown on SrTiO\textsubscript{3} (STO) substrates. Since then, the family of superconducting (SC) IL nickelates has significantly expanded in the thin film form including Sr:PrNiO\textsubscript{2},\cite{Osada(2020),Osada(2020)PRM} Ca:LaNiO\textsubscript{2},\cite{Zeng(2022)} Sr:LaNiO\textsubscript{2}\cite{Osada(2021)} and Eu:NdNiO\textsubscript{2}\cite{Wei(2023)SciAdv} compounds. To date, superconductivity is absent in bulk nickelates,\cite{Puphal(2021),QinLi(2020)} leading to early speculation regarding a possible interfacial character. This hypothesis now seems to be discarded after recent observations of superconductivity in nickelate thin films grown on an alternative substrate  (LaAlO\textsubscript{3})\textsubscript{0.3}(Sr\textsubscript{2}TaAlO\textsubscript{6})\textsubscript{0.7}(LSAT)\cite{KyuhoLee(2023),Ren_LSAT(comphysics)} as well as different interface terminations for SC IL films on STO substrates.\cite{Berit(2023), (Yang2023), Aravind(2024)}

Nickelates now constitute an entire new realm of research, with a number of outstanding questions needing to be addressed to clarify whether nickelates are true analogs of cuprates or a distinct family of unconventional superconductors.\cite{Lechermann(2020), Botana(2021), Chow(2022), Alvarez(2022), Kitatani(2023)} However, the synthesis of these compounds faces significant challenges that currently impede further progress of this nascent field.

One of the main difficulties lies in the intricate topotactic reduction process used to selectively remove only the oxygens at the apical sites of the initial parent perovskite phase and end up with the SC IL phase. It requires pushing Ni to the thermodynamically unstable valence state $Ni^{+1}$ to reach the desired $3d^{9}$ configuration, whereas the most common stable oxidation state is $Ni^{+2}$.\cite{Rice(1999)} In addition, the conventional reduction method employing CaH\textsubscript{2} as the reducing agent presents reproducibility problems between different research groups.\cite{Lee(2020), Chow(2022), Wang(2024)} The superconducting samples obtained by this method also lack of crystallinity at the top surface and seem vulnerable to re-oxidation upon exposure to air for a prolonged time.\cite{Osada(2020),Osada(2020)PRM,Zeng(2022),Osada(2021),Li(2019),LiYueying(2021),KyuhoLee(2023), Aravind(2023)} To address this issue, a protective STO capping layer is usually added,\cite{Daniele(2023)} which is hindering the study of the electronic structure or superconducting gap by using surface-sensitive techniques, such as angle-resolved photoemission spectroscopy (ARPES) and scanning tunneling microscopy (STM). It is therefore crucial for the advancement of the field to develop alternative reliable reduction methods that could overcome these obstacles and provide high quality superconducting nickelate samples.

Last year, an alternative reduction approach was proposed by Wei \emph{et al.}\cite{Wei(2023),Wei(2023)SciAdv} The perovskite parent phase is reduced \textit{in situ} by depositing an aluminum (Al) overlayer using molecular beam epitaxy (MBE). Furthermore, earlier this year, two independent groups also reported the use of \textit{in situ} atomic hydrogen bombardment to successfully reduce the perovskite phase into the SC infinite-layer phase.\cite{Kyle(2024)Nature,Kyle(2024)APL,Sun(2024)} However, these techniques are limited in their accessibility for many research groups, highlighting the need to find simpler \textit{ex situ}  or \textit{in situ} methods to reduce the nickelates.

In this work, we demonstrate an effective and simpler alternative route to synthesize high quality superconducting IL Pr\textsubscript{0.8}Sr\textsubscript{0.2}NiO\textsubscript{2} (PSNO\textsubscript{2}) thin films. We use a more accessible technique such as direct current (DC) magnetron sputtering to deposit a thin aluminum metal layer on top of the perovskite parent Pr\textsubscript{0.8}Sr\textsubscript{0.2}NiO\textsubscript{3} (PSNO\textsubscript{3}) thin films. The Al overlayer selectively pumps the apical oxygen atoms from the perovskite thin films through an efficient redox reaction and attains the complete transformation of the nickelates thin films into the SC IL phase. This approach is somewhat reminiscent of that used to generate two-dimensional electron gases (2DEGs) in SrTiO$_3$\cite{Rodel(2016), Vicente(2021)} and KTaO$_3$,\cite{Vicente(2021)Adv, Mallik(2022)} but here with a stronger reduction efficiency in optimized conditions.

We describe the whole optimization process of the Al deposition conditions as well as the comparison between SC samples reduced by Al deposition either \textit{in situ} or \textit{ex situ} (\textit{i.e.}, after having exposed the precursor film to the air). By using this method, we obtain high quality SC infinite-layer PSNO\textsubscript{2} thin films with a maximum superconductivity transition $T_{c}^{onset}$ of 17 K when the whole process is carried out \textit{in situ}. This new approach allows better control over the chemical transformation and improve sample reproducibility and quality. It also grants the possibility to perform \textit{in situ} reduction and characterization while preserving the surface cleanliness, which, combined with the top-down nature of the process, enables the use of surface-sensitive techniques. This new, much more accessible method, enriches the catalog of synthetic recipes available to date for obtaining infinite-layer nickelates, and is expected to encourage more research groups to synthesize these compounds, which could contribute to the physical understanding of the superconductivity in IL nickelates.

\section{Results and discussion}

Precursor Pr\textsubscript{0.8}Sr\textsubscript{0.2}NiO\textsubscript{3} perovskite thin films were grown by Pulsed Laser Deposition (PLD) on (001)-oriented STO substrates following previously optimized conditions described in the experimental section.\cite{Araceli(2024)} Figures \ref{fig:S1}-\ref{fig:S3} of the supporting information show a summary of the structural characterization of the as-grown perovskite thin films. \textit{In situ} Reflection High-Energy Electron Diffraction (RHEED) monitoring of the growth indicates a layer-by-layer growth of the films and a flat surface exhibiting step terraces structure. The films also exhibit clear pseudocubic perovskite 00\textit{l} peaks, with no additional features, indicating a high crystalline quality and single-phase perovskite PSNO\textsubscript{3} films. The out-of-plane lattice constant (\textit{c}) is $\approx$ 3.76 \AA, consistent with previous reports on this compound.\cite{Osada(2020),Osada(2020)PRM,Araceli(2024)} In addition, Reciprocal Space Map (RSM) collected around the asymmetric ($\bar{1}$03) peak of SrTiO$\textsubscript{3}$ confirms that the films are coherently strained to the substrate. The subsequent reduction of the precursor perovskite films is achieved by depositing an Al layer using DC magnetron sputtering technique. The Al deposition is carried out at specified temperature and is followed by a post-deposition annealing at the same temperature. The schematic of the process is shown in Figure \ref{fig:optimisation} a). During the reduction, the Al pumps oxygen from the parent perovskite nickelate film through an efficient redox reaction and gets oxidized following the chemical reaction:

\begin{equation}
	3PrNiO\textsubscript{3} + 2Al \rightarrow 3PrNiO\textsubscript{2} + Al\textsubscript{2}O\textsubscript{3}
\end{equation}

\subsection{Aluminum reduction optimization}

Firstly, we focused on systematically optimizing the deposition conditions of the Al layer. In the case of DC magnetron sputtering deposition, several parameters may play a crucial role in the success of the parent phase reduction. Notably, the reduction temperature, defined as the temperature at which we deposit the aluminum layer, the deposition rate and the post-annealing time. All these parameters have been tuned to optimally reduce the parent PSNO\textsubscript{3} film and achieve the complete transformation into the infinite-layer phase. It is also worth mentioning that we have empirically verified a negligible effect of the pre-annealing time before the Al deposition in the synthesis of the IL films. 

During the optimization of the parameters mentioned above, the Al deposition was consistently performed \textit{ex situ}. Thus, prior to any Al deposition, all the perovskite thin films were grown on $10\times10~mm^{2}$ STO substrates and removed from the PLD chamber to verify their crystalline quality and transport properties. Subsequently, each film was cut into four pieces of $5\times5~mm^{2}$ to test various Al deposition conditions. Technical details regarding the Al sputtering deposition are provided in the experimental section of this paper.

\begin{figure*}[!ht]
	\includegraphics[width=1.01\textwidth]{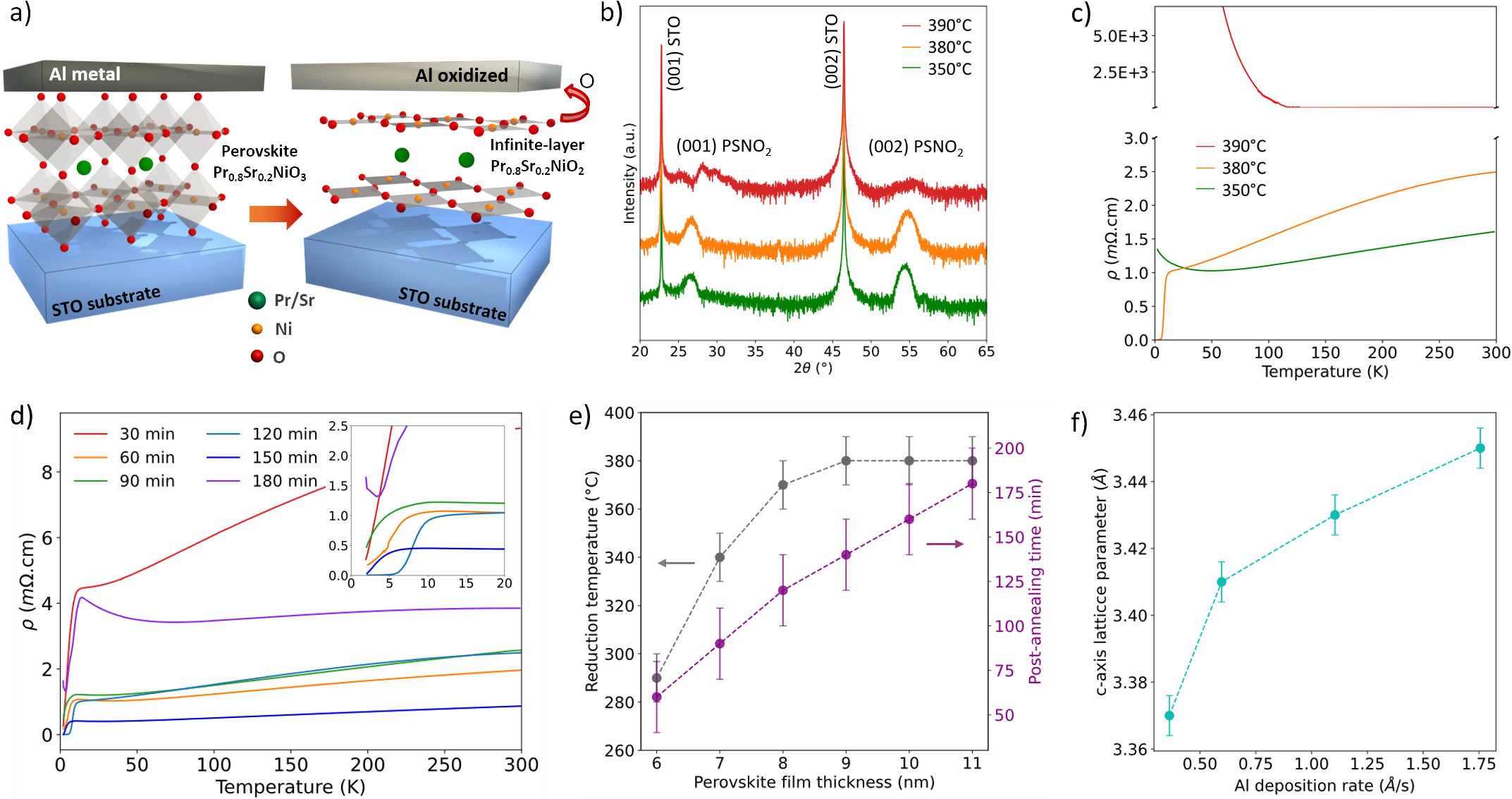}
	\caption{a) Schematic diagram of the Al reduction process in nickelates. First, an aluminum overlayer is sputtered onto the precursor PSNO\textsubscript{3} perovskite thin film at moderate temperatures (280 - 380°C). This is followed by a post-deposition annealing at the same temperature for a specified time (60 - 180 min). Al deposition temperature and post-annealing time are selected depending on the thickness of the precursor perovskite film. b) X-Ray Diffraction $\theta$ - 2$\theta$ symmetric scans of a 8-nm thick parent PSNO\textsubscript{3} thin film reduced under different Al deposition temperatures, 350, 380 and 390°C. The post-annealing time was fixed to 120 minutes in all the cases. c) Resistivity as a function of the temperature for the samples showed in panel b). d) Temperature dependence of resistivity as a function of the post-annealing time for a representative series of 8 nm-thick films when Al is deposited at 380°C. The inset shows the same data near the superconducting transition. e) Evolution of the Al deposition temperature (grey circles) and post-annealing time (purple circles) as a function of the perovskite PSNO\textsubscript{3} film thickness. The Al thickness was set at 3.5 nm. f) Evolution of the c-axis lattice parameter (Å) of the thin film as a function of the aluminum deposition rate in Å/s. The deposition rate optimization was performed on a 6 nm-thick sample (10x10 mm$^{2}$, later cut in four pieces for the experiments), using 2.5 nm Al thickness, at a fixed reduction temperature of 290°C and a post-annealing time of 60 mins.}
	\label{fig:optimisation}
\end{figure*}

We initially optimized the Al deposition temperature (hereafter, reduction temperature) for an 8 nm thick film. The aluminum thickness used for these initial experiments was $\approx$ 3.5 nm, following the findings of previous work on Al reduction by MBE.\cite{Wei(2023)} The optimal reduction temperature was determined by combining X-ray Diffraction (XRD) and transport measurements to monitor the crystalline quality and the presence/absence of a superconducting transition in the thin films. Figure \ref{fig:optimisation} b) and c) depict the diffractograms and the temperature-dependent resistivity curves of a series of 8 nm thick perovskite films with Al deposition performed at various temperatures. When the Al deposition is performed at 350°C, the film partially transforms into the IL phase. Although it shows good crystalline quality, with intense (001) and (002) PSNO\textsubscript{2} film peaks and 2$\theta$$=$54.8° ($\textit{c}$ $\approx$ 3.34\AA), it does not undergo a superconducting transition and exhibits  insulating nature at low temperatures. It is important to note that this value for the out-of plane lattice parameter (\textit{c}) aligns with those reported in the literature for the stabilized SC IL PSNO\textsubscript{2} phase.\cite{Osada(2020),Osada(2020)PRM,Araceli(2024)} As reduction temperature increases to 380°C, a complete transformation into the IL phase is evidenced by a minimal difference in the crystalline quality of the film (2$\theta$$=$55°, c $\approx$ 3.34\AA), but mainly by a superconducting transition that reaches a zero-resistance state. However, when Al deposition occurs at temperatures above 380°C, the film suffers a loss of crystallinity or may even decompose, as can be inferred by the significant reduction in intensity of both (001) and (002) PSNO\textsubscript{2} film peaks. This over-reduced sample exhibit clear insulating behavior at low temperatures. These results highlight the importance of using not only XRD but also transport measurements to confirm the correct stabilization of the superconducting phase in nickelates.

Once Al reduction temperature was set at 380°C for a film of 8 nm thickness, we explored the effect of the post-annealing time, defined as the period during which the sample is maintained at the deposition temperature after the completion of Al deposition. This post-annealing time controls the oxygen diffusion from the bottom layer of the film, adjacent to the substrate interface, to the top surface where oxygen oxidizes the aluminum layer. As demonstrated by STEM analysis conducted Wei \textit{et al.},\cite{Wei(2023)} the topotactic reduction driven by aluminum deposition proceeds in a top-down manner, which makes optimization of the post annealing time essential. In Figure \ref{fig:optimisation} d), we present temperature-dependent resistivity measurements for a series of 8 nm thick samples prepared with different post-annealing time, while all other parameters were kept constant. For short post-annealing times ranging from from 30 to 90 minutes, the films are not fully reduced showing partial superconducting transitions without reaching zero resistance state. In contrast, for an optimal post-annealing time of 120 minutes, a complete superconducting transition and zero resistance state are observed. The onset transition temperature ($T_{c}^{onset}$), defined from now on as the intersection of linear extrapolations from the normal state and the superconducting transition regions, is $T_{c}^{onset}$$=$10 K. Exceeding this optimal time deteriorates the superconducting properties of the film, with a significant decrease of the $T_{c}^{onset}$ (at 150 min), or even results in an over-reduced film (at 180 min). The latter causes a marked upturn starting at 50 K, followed by a partial superconducting transition at lower temperature, but without reaching zero resistance. In addition to the substantial resistivity changes  depending on the post-annealing time, XRD measurements show a notable improvement in film crystallinity as the post-annealing time increases from 10 min to 60 min (see supporting information, Figure \ref{fig:S4}). For longer post-annealing times, the improvement in XRD data is minimal and the c-axis lattice parameter of the thin film varies little between 120 to 180 min. Therefore, superconducting IL samples can be achieved within a relatively narrow post-annealing time window of around 120 minutes for the 8 nm thick film presented here. This result confirms the fundamental role of the post-annealing time in achieving a complete reduction of the entire thickness of the perovskite film.

We carried out similar optimization of reduction temperature and post-annealing time for different perovskite film thickness between 6 and 11 nm, as is shown in Figure \ref{fig:optimisation} e). We could not investigate films thicker than 11 nm due to the reported poorer quality of the parent perovskite films, characterized by the appearance of secondary phases.\cite{Li(2019)} At first glance, one can notice that the two parameters are closely correlated: as film thickness increases, it is necessary to simultaneously raise both the deposition temperature and the post-annealing time to achieve complete reduction of the nickelate thin films. However, the dependence of both parameters with the perovskite film thickness differs. While the post-annealing time exhibits clear linearity, the reduction temperature increases rapidly for low thicknesses (6 - 8 nm) and then remains constant for films thicknesses between 9 - 11 nm.

So far we have used 3.5 nm Al thickness to reduce the samples, but it is also important to determine the appropriate Al thickness needed to completely reduce the perovskite PSNO\textsubscript{3} into the IL PSNO\textsubscript{2} phase, while avoiding any potential contribution from an Al metal overlayer in the transport measurement. In addition, the aluminum oxide overlayer formed during reduction also efficiently protects the nickelate from re-oxidation, serving as both a reducing agent and as a protective layer. Thus, we fine-tuned the Al thickness to ensure the complete oxidation of the overlayer into Al\textsubscript{2}O\textsubscript{3}. For this purpose, we \textit{in situ} monitored the evolution of the X-Ray photoelectron spectroscopy (XPS) spectra of the core level spectra of Al 2p corresponding to the Al metal ($Al^{0}$) and oxidized Al ($Al^{3+}$) (see Figure \ref{fig:S5} in the supporting information) after each Al deposition process is completed. The Al layer thickness is then reduced until no trace of Al metal is observed. Following this optimization, the Al thickness was reduced to 2.5 - 2.8 nm for the subsequent experiments.

We finally focused on determining the optimal aluminum deposition rate by varying it from 0.363 \AA/s to 2.357 \AA/s. To elucidate the optimal growth rate, we use complementary X-ray diffraction (XRD) measurements to monitor the evolution of the c-axis lattice parameter as a function of the deposition rate. As shown in Figure \ref{fig:optimisation} f), the trend is clear: the lowest deposition rate enhances the reduction process, resulting in the smallest c-axis lattice parameter, which is approximately 3.37 \AA. Therefore, the aluminum deposition rate was fixed at 0.363 \AA/s (or $\approx$ 22 \AA/min) for the subsequent experiments. Note that this deposition rate is significantly faster than the rate normally used during the Al deposition by other techniques such as MBE ($\approx$ 0.5 layer/min $\approx$ 0.54 \AA/min).\cite{Wei(2023)}

\subsection{Differences between \textit{ex situ} and \textit{in situ} reduction}

\begin{figure*}[!ht]
	\includegraphics[width=1\textwidth]{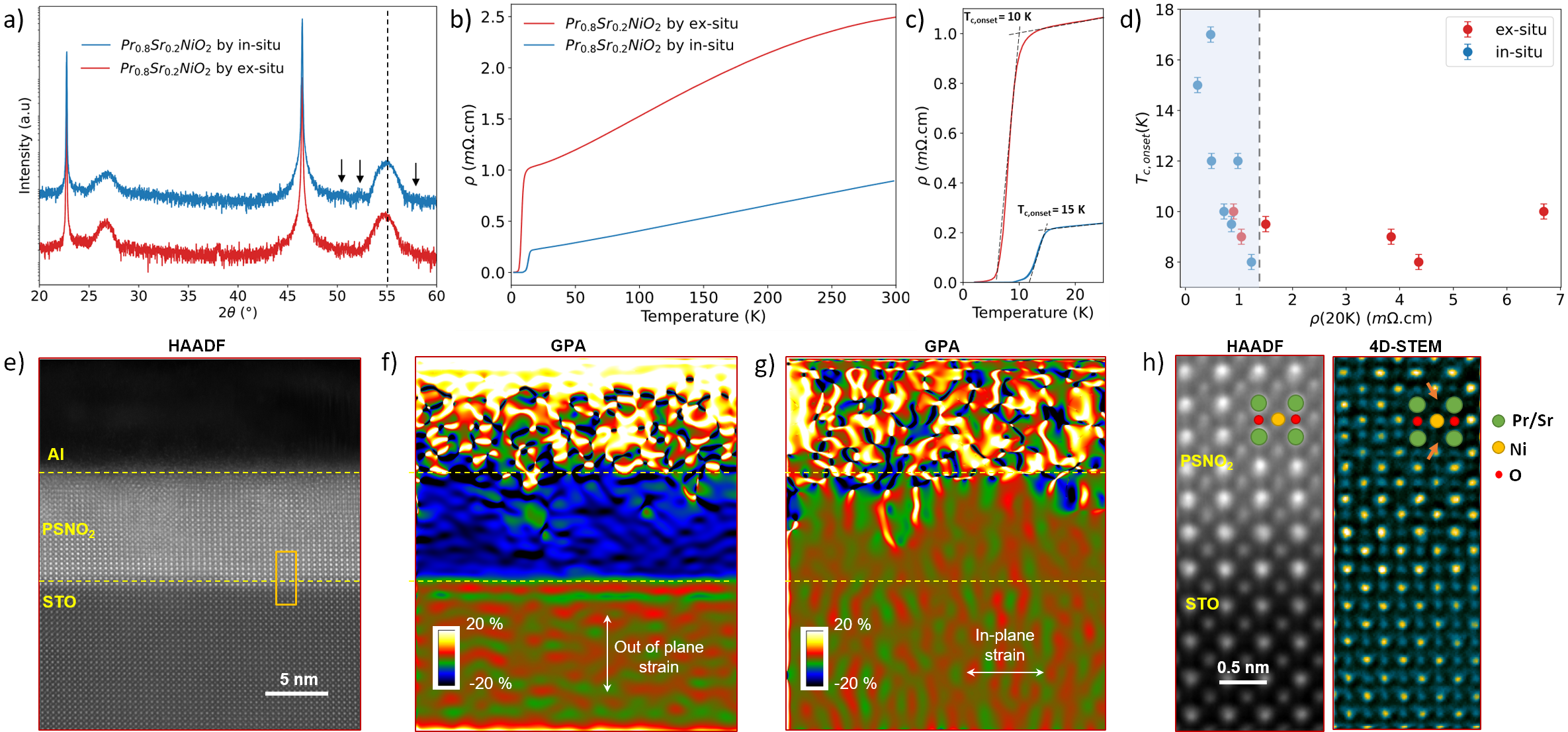}
	\caption{a) X-Ray Diffraction $\theta$ - 2$\theta$ scans of aluminum reduced Pr\textsubscript{0.8}Sr\textsubscript{0.2}NiO\textsubscript{2} samples, one exposed to air before aluminum deposition (\textit{ex situ}, red), and another reduced without air exposure (\textit{in situ}, blue). b) Resistivity comparison as a function of the temperature for infinite-layer phase after \textit{ex situ} aluminum reduction (red line), showing $T_{c}^{onset}$ = 10 K and $T_{c}^{zero}$ = 4 K, and after \textit{in situ} aluminum reduction, showing $T_{c}^{onset}$ = 15 K and $T_{c}^{zero}$ = 9.5 K. c) Enlarged view of the $\rho$ versus T curves from panel b) in the temperature range of 2 to 25 K (around the superconducting transition). $T_{c}^{onset}$ is defined as the intersection of the linear extrapolations of the normal state and the superconducting transition regions. d) Critical temperature onset ($T_{c}^{onset}$) as a function of the resistivity ($\rho$) at 20 K for \textit{in situ} and \textit{ex situ} samples, respectively. Blue shaded area is a guide to the eye indicating the most common values for \textit{in situ} reduced samples.  e) HAAF-STEM image of a Pr\textsubscript{0.8}Sr\textsubscript{0.2}NiO\textsubscript{2} film reduced via \textit{ex situ} Al deposition. The nominal thickness of the initial perovskite film was 8 nm with 3.5 nm of Al used for the reduction. f,g) Geometrical phase analysis the of STEM image in panel e) showing the out-of-plane (f) and in-plane (g) lattice contraction relative to the SrTiO\textsubscript{3} substrate. h) (left panel) Atomic-resolution HAAF-STEM image from the region near the bottom interface, marked in panel e) (orange rectangle), and (right panel) corresponding 4D-STEM dCOM image showing the absence of oxygen atoms at the apical sites (indicated by orange arrows). In both images, green circles represent Pr/Sr atom sites, orange circles represent Ni sites and red circles represent O sites. }
	\label{fig:exsitu_vs_insitu}
\end{figure*}

After optimizing the parameters for Al reduction, we analyzed the main differences observed in samples reduced via \textit{ex situ} and \textit{in situ} processes. \textit{In situ} experiments were performed on PSNO\textsubscript{3} thin films grown $5\times5~mm^{2}$ (001) SrTiO\textsubscript{3}  substrates, which were directly transferred to the sputtering chamber for Al deposition in ultra-high vacuum without any exposure to air.

Firstly, as illustrated by the XRD measurements of Figure \ref{fig:exsitu_vs_insitu} a), minimal changes are observed in the crystallinity of the samples reduced by both methods. \textit{In situ} reduced films seems to exhibit a slightly higher crystalline quality, evidenced by the tiny Laue oscillations around the (002) peak of the film (indicated by arrows in the figure). Further, the (002) peak of the \textit{in situ} reduced film is shifted towards higher angles, indicating a minor decrease in its out-of-plane lattice parameter (c $\approx$ 3.34 \AA). However, an improvement of \textit{in situ} reduced samples compared to \textit{ex situ} ones is more evident in the transport properties, as can be seen in Figure \ref{fig:exsitu_vs_insitu} b). Although both \textit{in situ} and \textit{ex situ} samples reach a zero-resistance state, a substantial decrease in overall resistivity as well as an enhancement of $T_{c}^{onset}$ can be observed in the \textit{in situ} reduced film (see Figure \ref{fig:exsitu_vs_insitu} c)). The absolute value of resistivity for the \textit{in situ} reduced sample is $\rho$ (20K) $\approx$ 0.25m$\Omega$.cm, which matches the reported values for SC PSNO\textsubscript{2} films on STO substrate\cite{Osada(2020),Osada(2020)PRM} and represents a quarter of the value obtained from the \textit{ex situ} reduced sample. Remarkably, the resistivity of the \textit{in situ} reduced film clearly exhibits a $T$-linear behavior with only a mild upturn in resistivity just before the SC onset, whereas the \textit{ex situ} film does not show a $T$-linear behavior and is followed by a marked resistivity upturn at low temperatures (see linear fittings on supporting information, Figure \ref{fig:S6}). Consistent with previous reports,\cite{KyuhoLee(2023)} this observation suggests a notable decrease in disorder in the \textit{in situ} reduced films compared to those exposed to the air during the intermediate step between PLD growth and Al sputtering deposition. The slope of the linear fitting for \textit{in situ} reduced film is roughly 24$\times$10$^{-4}$ m$\Omega$.cm.K$^{-1}$ (with variations among samples), comparable to the values found in other SC IL nickelates grown on STO, such as (Nd,Sr)NiO\textsubscript{2}.\cite{KyuhoLee(2023)} Other families of unconventional superconductors, such as cuprates,\cite{Cooper(2009), Legros(2019)} also exhibit a linear dependence of resistivity, although its origin is not clearly understood.\cite{Bruin(2013)}

Furthermore, the width of the SC transition (defined as 90$\%$-10$\%$ of the normal state at 20 K) also slightly decreases for the \textit{in situ} reduced samples ($\Delta T_{c}\approx 2-3~K$) compared to the \textit{ex situ} ones ($\Delta T_{c}\approx 3-4~K$), indicating an improvement in the homogeneity of the superconductivity in \textit{in situ} reduced samples. Previously reported values of the superconducting transition width for samples reduced using conventional CaH\textsubscript{2} reduction\cite{Araceli(2024)} ($\Delta T_{c} \approx 3~K$) are comparable to our \textit{in situ} Al reduced samples.

Figure \ref{fig:exsitu_vs_insitu} d) plots the $T_{c}$ values as a function of resistivity ($\rho$) at 20 K for several samples reduced by \textit{ex situ} and \textit{in situ} Al deposition. The \textit{ex situ} reduced samples exhibit a wide range in the resistivity values (1 to 7 m$\Omega$.cm), indicating limited control over the disorder in the samples, while the $T_{c}$ values always remain below 10 K. In contrast, the \textit{in situ} Al reduced samples display a much more reproducible $\rho$ (20 K) value, consistently falling below 1.2 m$\Omega$.cm. However, much more variability is observed in the $T_{c}$ values for \textit{in situ} reduced samples, reaching a maximum $T_{c}$  of 17 K for the best sample obtained by this method. This further supports the idea of an improvement in synthesis regarding disorder for the \textit{in situ} Al reduced samples. Nonetheless, we notice some variation in the transition temperature from sample to sample, potentially related to the crystallinity of the parent perovskite phase or local differences in oxygen stoichiometry. Recent reports on \textit{in situ} hydrogen reduction of nickelates show similar variability in $T_{c}$ values but the underlying cause remains unclear\cite{Sun(2024)}. This aspect will be studied in more detail and, if possible, improved in future studies.

The observed reduction in disorder in \textit{in situ} reduced samples compared to \textit{ex situ} is not fully understood, but it is likely related to the quality of the top interface with the Al. Indeed, scanning transmission microscopy (STEM) experiments performed in a \textit{ex situ} Al reduced sample provide further insight into this aspect. The cross sectional high-angle dark-field (HAADF) image in Figure \ref{fig:exsitu_vs_insitu} e) shows a clean infinite-layer structure, particularly near the substrate, but evidences more defects in the upper region of the IL film and an amorphous interface with the $AlO_{x}$ overlayer. Geometrical phase analysis (GPA) applied to that image reveals few vertical Ruddlesden-Popper defects and strain inhomogeneities in both out-of-plane (Figure \ref{fig:exsitu_vs_insitu} f)) and in-plane strain (Figure \ref{fig:exsitu_vs_insitu} g), which are mainly originated from the top interface. However, the bottom interface exhibits a properly stabilized IL phase after Al reduction. In this regard, Figure \ref{fig:exsitu_vs_insitu} h) depicts an HAADF image (left panel) taken at the bottom interface, and the corresponding 4D-STEM divergence of center of mass (dCOM) analysis (right panel) confirms the absence of apical oxygen atoms (indicated by arrows), with a resulting structure of alternating Pr/Sr and NiO\textsubscript{2} planes (see supporting information, Figure \ref{fig:S7} for detailed analysis).

In \textit{ex situ} reduced samples, the surface of the perovskite film is exposed to air prior to Al deposition, and XPS measurements (not shown here) revealed a significant amount of adventitious carbon adhered to this surface (that is almost negligible in \textit{in situ} reduced samples). This contamination could hinder or cause inhomogeneous oxygen diffusion towards the Al layer during the subsequent reduction process and could explain the poorer crystalline quality of the film near the upper interface and the overall lower quality of \textit{ex situ} reduced samples. Developing a cleaning protocol (possibly involving an oxygen or argon plasma) to obtain high-quality samples via the \textit{ex situ} process is necessary.\cite{Rath(2024)} Although this topic is beyond the scope of this work, it represents a promising direction for future research.

There are few additional observations regarding \textit{ex situ} and \textit{in situ} reduction worth discussing. First, while superconducting IL samples are achieved using both processes, the \textit{in situ} reduction reliably produces SC films with zero-resistance state and superior quality in terms of crystallinity and transport properties (e.g. enhanced crystallinity, $T$-linear behavior, higher T\textsubscript{c} value and sharper transitions). In contrast, the \textit{ex situ} process presents notable sample-to-sample variation, often resulting in samples with only a partial superconducting transitions without reaching the zero resistance state, indicating a less consistent outcome. Second, note that for \textit{in situ} reduced samples, there is no apparent hydrogen source as the process is carried out entirely in vacuum, yet we obtain high-quality superconducting IL samples. This aligns with recent studies indicating that extensive hydrogen incorporation is not a key ingredient in achieving superconductivity in IL nickelates.\cite{Balakrishnan(2024), Zeng(2024), Gonzalez(2024)} Lastly, even in \textit{in situ} reduced superconducting samples, some inhomogeneity is observed within the $5\times5~mm^{2}$ samples, with slight variations in $T_{c}$ and resistivity depending on the direction of the transport measurements performed in van der Pauw configuration (see supporting information, Figure \ref{fig:S8}). We could not find reports indicating whether this issue is commonly encountered in superconducting samples produced by other research groups. However, this inhomogeneity could be significant for experiments requiring consistent pieces from the same sample and suggests that there is still room for improvement in achieving fully homogeneous SC samples.

\subsection{Characterization of an \textit{in situ} Al reduced superconducting infinite-layer thin film}

\begin{figure}[!ht]
	\includegraphics[width=0.48\textwidth]{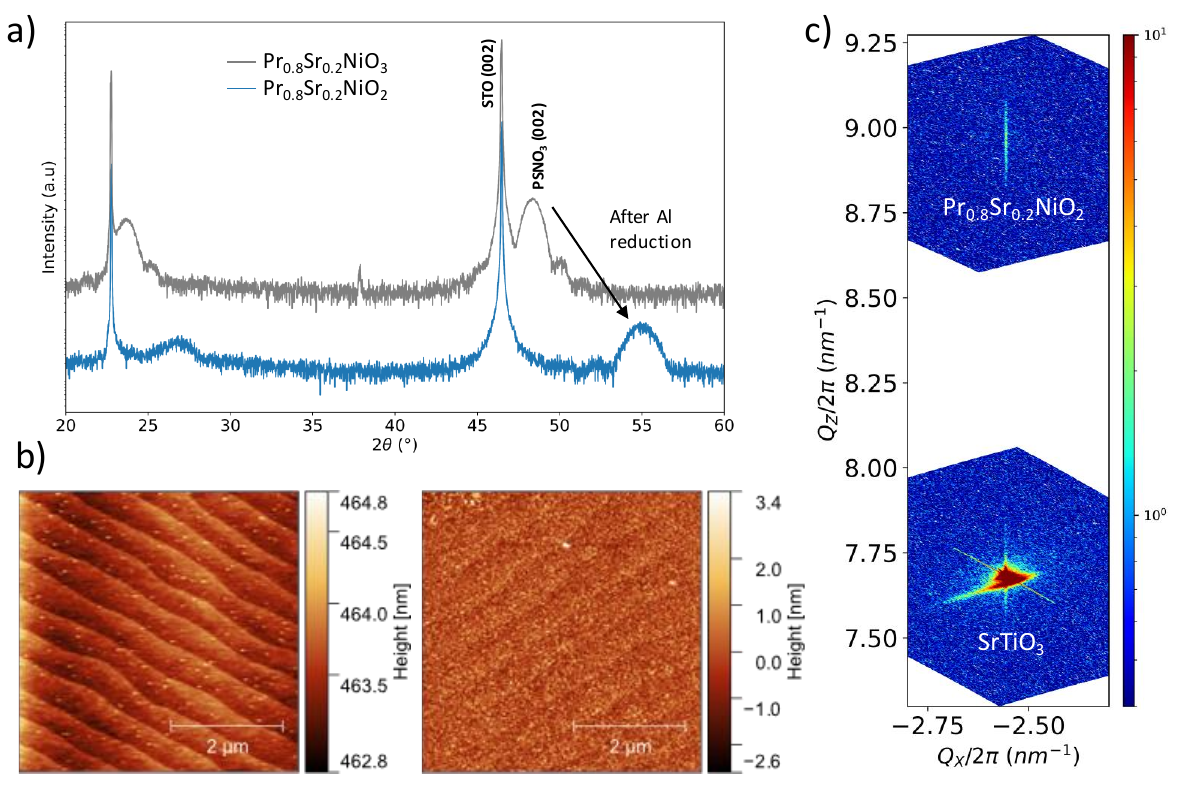}
	\caption{a) X-Ray Diffraction $\theta$ - 2$\theta$ scans of a 8-nm thick parent Pr\textsubscript{0.8}Sr\textsubscript{0.2}NiO\textsubscript{3} thin film (grey) and \textit{in situ} aluminum reduced sample (Pr\textsubscript{0.8}Sr\textsubscript{0.2}NiO\textsubscript{2} film) (blue). b) (left panel) AFM topography images of the parent perovskite phase (average steps height $\approx$ 0.595 nm, average steps width $\approx$ 324 nm) and (right panel) the \textit{in situ} reduced sample after aluminum deposition by DC magnetron sputtering (average roughness $\approx$ 0.517 nm). c) High resolution RSM around the ($\bar{1}$03) asymmetric reflection of an \textit{in situ} reduced infinite-layer nickelate thin film on SrTiO\textsubscript{3} substrate, indices are taken with respect to the pseudocubic unit cell.}
	\label{fig:growth}
\end{figure}

Having demonstrated an enhancement in the crystalline and transport properties of the \textit{in situ} reduced samples, we now focus on providing a more exhaustive characterization of one of those SC IL samples. Figure \ref{fig:growth} a) shows the structural characterization by XRD of the parent nickelate thin film before and after \textit{in situ} Al deposition. An as grown 8-nm thick PSNO\textsubscript{3} thin film presents the (002) peak at $\approx$ 48.34° (in grey) (c$=$3.76\AA), consistent with previously reported values.\cite{Osada(2020),Osada(2020)PRM} After the deposition of 2.5 nm of Al via DC magnetron sputtering, the (002) film peak (in blue) shifts to higher 2$\theta$ angle, indicating a decrease of the c-axis lattice parameter (c = 3.352 \AA, 2$\theta$ = 54.74°) and the complete transformation of the perovskite into the infinite-layer phase. The aluminum deposition was performed under the previously optimized conditions: a temperature of 380°C, a post-annealing time of 120 minutes and the deposition rate fixed at 0.363 \AA/s.

\begin{figure*}[!ht]
	\includegraphics[width=0.95\textwidth]{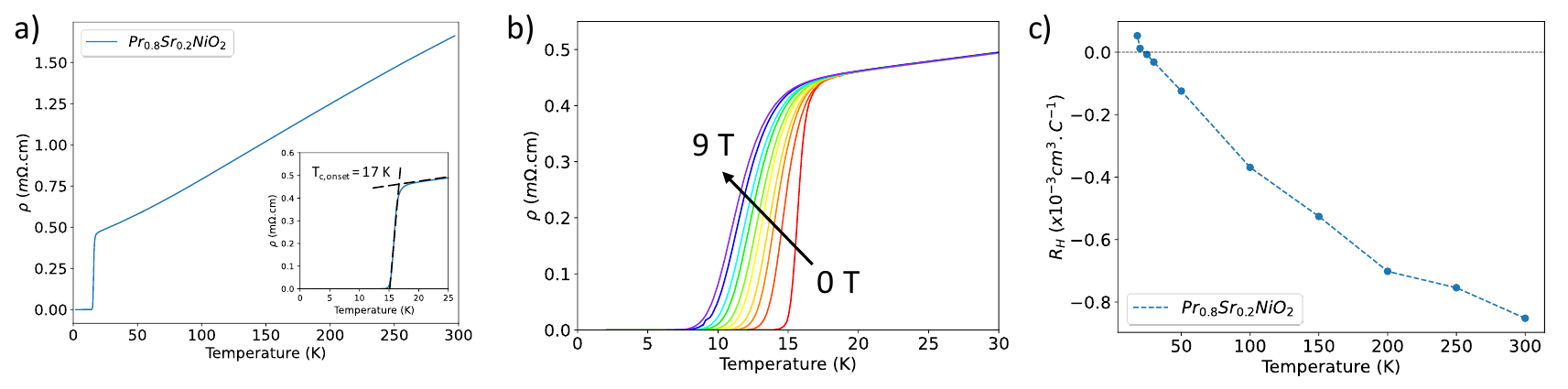}
	\caption{a) Resistivity as a function of temperature for the superconducting infinite-layer phase obtained after \textit{in situ} aluminum reduction, with a $T_{c}^{onset}$ = 17 K and $T_{c}^{zero}$ = 15 K. The inset shows an expanded view around the superconducting transition. b) Temperature-dependent resistivity measurements of an aluminum reduced Pr\textsubscript{0.8}Sr\textsubscript{0.2}NiO\textsubscript{2} sample around the $T_{c}^{onset}$ for different externally applied out-of-plane magnetic fields up to 9 T. c) Normal state Hall coefficient (R\textsubscript{H}) as a function of temperature for \textit{in situ} Al reduced PSNO$\textsubscript{2}$ thin films.}
	\label{fig:electrical}
\end{figure*}

The surface topography of the as-grown and \textit{in situ} Al reduced sample were characterized by atomic force microscopy (AFM), shown in Figure \ref{fig:growth} b). After aluminum deposition, the surface remains well-preserved and smooth, with a root mean square roughness of 0.5 nm, indicating that the Al\textsubscript{2}O\textsubscript{3} overlayer is uniform. Additionally, the surface exhibits step terraces with an average step height of $\approx$ 0.595 nm (one and a half times the thickness of a STO unit cell). The reciprocal space map (RSM) of the SC infinite-layer phase, shown in Figure \ref{fig:growth} c), further demonstrates that the film remains fully strained to the SrTiO\textsubscript{3} substrate, after reduction by aluminum deposition.

Detailed electrical transport measurements were performed for an \textit{in situ} reduced thin film, as shown in Figure \ref{fig:electrical} a). The SC infinite-layer phase displays a superconducting transition with a $T_{c}^{onset} \approx$ 17 K, reaching zero resistance with $T_{c}^{zero}$ $\approx$ 15 K. These values are comparable to, or in some cases even higher than, those previously reported using traditional CaH\textsubscript{2} reduction for the PSNO\textsubscript{2} phase.\cite{Osada(2020),Osada(2020)PRM, Wang_highquality(2022)Natcom} The $T$-linear slope of the IL nickelate is roughly 43.9 $\times 10^{-4}~m\Omega.cm.K^{-1}$, and when extrapolated to zero, it intercepts at 0.359 m$\Omega$.cm. As previously mentioned, these values are comparable to the current state of the art of PSNO\textsubscript{2} and NdSrNiO\textsubscript{2} thin films grown on STO substrate by other groups of research.\cite{Osada(2020), Osada(2020)PRM, Li(2019)} Optimally doped NdSrNiO\textsubscript{2} films grown on LSAT substrate have reported slopes of roughly 11 $\times 10^{-4}~m\Omega.cm.K^{-1}$,\cite{KyuhoLee(2023)} attributed to improved crystallinity and reduced disorder of the films owning to a favorable epitaxial mismatch for the perovskite and the infinite-layer phases on LSAT substrate compared to STO.

Furthermore, Figure \ref{fig:electrical} b) shows the temperature-dependent resistivity when an out-of-plane magnetic field is applied, up to 9 T. We observe that the SC transition gradually shifts to lower temperatures, and the transition width broadens with increasing magnetic field. However, the superconductivity transition still remains at the highest magnetic field we can apply (9 T), in line with same type of measurements in other SC IL nickelates systems, such as Nd\textsubscript{1-x}Sr\textsubscript{x}NiO\textsubscript{2},\cite{Li(2019),Ding(2022)} Nd\textsubscript{1-x}Eu\textsubscript{x}NiO\textsubscript{2}\cite{Wei(2023)SciAdv} or La\textsubscript{1-x}Sr\textsubscript{x}NiO\textsubscript{2}.\cite{Osada(2021)}

The normal state Hall coefficient R\textsubscript{H} as a function of the temperature is presented in Figure \ref{fig:electrical} c). As observed in other nickelates compounds, R\textsubscript{H} remains negative at temperatures above $T_{c}$ but increases with decreasing temperature, changing sign from negative to positive while traversing the superconducting transition, specifically at 20 K (see Figure \ref{fig:S9} in supporting information). This suggests a complex Fermi surface with a mixed carrier contributions from both electrons and holes.\cite{Osada(2020)PRM,Araceli(2024),Li(2019),KyuhoLee(2023),Ariando,Ding(2022)} This change of sign observed in our 20$\%$ doped PSNO\textsubscript{2} thin films is consistent with what is reported for the same compound when the doping level varies from 12 and 32$\%$,\cite{Osada(2020)PRM} which are the superconducting compositions for PSNO\textsubscript{2} phase.

\section{Conclusion}
We have introduced a simple and accessible method for synthesizing superconducting infinite-layer nickelate Pr\textsubscript{0.8}Sr\textsubscript{0.2}NiO\textsubscript{2} thin films using aluminum sputtering deposition that can be performed either \textit{in situ} or after exposure of the perovskite film to air (\textit{ex situ}). All parameters involved in the Al sputtering deposition were systematically optimized, revealing that lower deposition rates enhance the reduction process, while the temperature of reduction and post-annealing time need to be adjusted based on the thickness of the parent perovskite phase. Detailed characterization of the resulting samples demonstrates that \textit{in situ} Al reduction reproducibly yields IL Pr\textsubscript{0.8}Sr\textsubscript{0.2}NiO\textsubscript{2} thin films with enhanced crystallinity and improved transport properties (T-linear behavior of resistivity, higher superconducting transition temperatures and sharper SC transition) compared to \textit{ex situ} reduced films.

Topotactic reduction of nickelates using Al sputtering deposition can be performed entirely \textit{in situ} while preserving an atomically flat surface and proceeds in a top-down manner, which provides an opportunity for surface-sensitive characterizations. This easily accessible \textit{ex situ}/\textit{in situ} approach also facilitates the synthesis of samples with quality comparable to those obtained through the reduction using CaH\textsubscript{2} as an agent, making it a promising alternative to such complex traditional method. More importantly, it offers a new pathway to expand the number of independent research groups capable of producing high quality superconducting infinite-layer nickelates samples, which could potentially boost experimental research aimed at understanding superconductivity in infinite-layer nickelates. 

\section{Experimental Section}
\textbf{\textit{SrTiO\textsubscript{3} substrates preparation and synthesis of perovskite thin films by PLD:}} Prior to growth, the (001) SrTiO\textsubscript{3} (SurfaceNet) substrates were etched in hydrofluoric acid (HF) solution and annealed at 1000°C in an oxygen atmosphere for 3 hours to obtain TiO\textsubscript{2} surface termination. To optimize the conditions for aluminum deposition, PSNO\textsubscript{3} thin films with  thicknesses ranging from 6 to 11 nm were grown on $10\times10~mm^{2}$ SrTiO\textsubscript{3} (001) substrate by pulsed laser deposition (PLD), which were then cut in four pieces of the same size.
For \textit{in situ} reduction, PSNO\textsubscript{3} thin films with a thickness of 8 nm (20-22 unit cells) were grown on $5\times5~mm^{2}$ SrTiO\textsubscript{3} (001) substrate.

Perovskite nickelate thin films were grown by PLD using a 248 nm KrF excimer laser. The films were deposited at a substrate temperature of 640°C and an oxygen partial pressure of 0.33 mbar, using an energy fluence of 1.6 $J/cm^{-2}$ (laser spot size $1\times2~mm^{2}$). After the growth, the film was cooled down at a rate of 10°C/min under the growth pressure. The growth of the films was monitored \textit{in situ} by a reflection high-energy electron diffraction (RHEED). More details on sample preparation and optimization can be found in ref.\cite{Araceli(2024)}

\textbf{\textit{\textit{ex situ} and \textit{in situ} Al reduction:}} When the process was carried out \textit{ex situ}, the films were removed from vacuum prior to Al deposition. In contrast, in the \textit{in situ} process, the films were directly transferred from the PLD chamber to the sputtering chamber without being expose to air. The deposition of the aluminum metal layer was carried out inside sputtering chamber (PLASSYS), at a pressure of $6.4\times10^{-4}$ mbar. The argon flow was set to 5.2 sccm, the current to 15 mA, power to 5 W and voltage to 320 V. Before any aluminum metal deposition the target was pre-sputtered for 10 minutes to remove any potential oxidized layer. The Al deposition to reduce the thin films was performed at sample temperatures ranging from 270 to 380°C in continuous mode. A post-annealing step at the deposition temperature was performed after the Al deposition in all the cases, varying the time from 30 to 180 minutes. The heating and cooling rates were set to 10°C/min.

\textbf{\textit{Structural and transport characterization}}: The structural quality and thickness of the thin films were characterized by X-ray diffraction (XRD) and X-ray reflectivity by using a X-ray diffractometer Empyrean (Malvern Panalytical).  The thickness was determined by fitting the XRR curves using the X'Pert Reflectivity software. The reciprocal space maps were obtained by using a rotating anode high-resolution X-ray diffractometer from Smartlab-Rigaku with a Cu ($K_{\alpha}$) = 1.5406 \AA~for both diffractometers. The surface topography of the samples was acquired with a Bruker Atomic Force Microscope in non-contact mode. Electrical transport measurements were performed using a Dynacool System (Quantum Design), with bondings done using gold or aluminum wires in van der Pauw geometry.

\textbf{\textit{Scanning transmission electron microscopy (STEM)}}: The cross-sectional lamellae for Transmission Electron Microscopy were prepared using a Focused Ion Beam (FIB) technique at Centre de Nanosciences et de Nanotechnologies (C2N), University Paris-Saclay, France. Prior to FIB lamellae preparation, around 20-30 nm of amorphous carbon was deposited on top of the samples for protection. The High-angle annular darkfield (HAADF) imaging and 4D-STEM was carried out in a NION UltraSTEM 200 C3/C5-corrected scanning transmission electron microscope. The experiments were done at 200 keV with a probe current of $\approx$ 12 pA and convergence semi-angles of 30 mrad. A MerlinEM (Quantum Detectors Ltd) in a 4 × 1 configuration (1024 x 256) had been installed on a Gatan ENFINA spectrometer mounted on the microscope\cite{Tence(2020)}. For 4D-STEM, the electron energy loss spectroscopy (EELS) spectrometer was set into non-energy dispersive trajectories and 6-bit detector mode that gave a diffraction pattern with a good signal to noise ratio without compromising much on the scanning speed was used. The geometrical phase analysis (GPA)\cite{Kilaas(1998)} had been done choosing the STO substrate with 3.91 \AA, as a reference parameter. The lattice parameters of the PSNO\textsubscript{2} were estimated by averaging the GPA maps over square areas of = 25 (in-plane) × 10 (out-of-plane) nm giving a strain accuracy determination better than 1\%, that is, better than 0.04 \AA for the lattice parameters. Such an approach has been previously employed to accurately determine the c-axis variation in an apical oxygen ordered nickelate thin-film on an STO substrate.\cite{Aravind(2023)}

\begin{acknowledgments}
	The authors thank A. Gutiérrez-Llorente for her help in the initial optimization of the perovskite nickelate thin films by PLD. We also acknowledge the support by Xiaoyan Li in the STEM experiments and V. Humbert for his help with the electrical characterization in a cryostat from Cryogenic and scientific discussion. We thank L. Matera for his help in the development of the Python codes and R. Tomar for the scientific discussion on transport data. D. Z. acknowledges financial support from École Doctoral 564 Physique en Ile de France (EDPIF) and Université Paris-Saclay. This work was also supported by the framework of the joint ANR-RGC ImagingQM project (RGC, A-CityU102/23; ANR, ANR-23-CE42-0027).  \\
	\\
	\textbf{Competing interests:} The authors declare that they have no competing interests.\\
	\\
	\textbf{Data availability:} The data that support the findings of this study are available from the corresponding author upon reasonable request.\\
		
	\centerline{\textbf{Author Contributions}}
	D.Z synthesized the samples (PLD growth), performed the experiments (aluminum reduction in sputtering chamber, XRD, XRR, electrical transport measurements) and analyzed the data. L.M.V.A. helped with the DC sputtering deposition process elaboration and data discussion. A.R and A.G performed the STEM experiments and the data analysis. M.B discussed the data and provided the necessary equipments and funding. L.I conceived and supervised the project, assisted in the experimental measurements (RSM, XRD, XRR), substrates treatment and data discussion. The manuscript was prepared by D.Z. and L.I. with input from all authors.
		
	\end{acknowledgments}
	
	\bibliographystyle{ieeetr}
	\bibliography{Biblio}

	\newpage
	\onecolumngrid
	\appendix
	
	\section*{Supplementary Information}

	\setcounter{figure}{0}
	\makeatletter 
	\renewcommand{\thefigure}{S\@arabic\c@figure}
	\makeatother
		
	\begin{figure*}[ht]
		\includegraphics[keepaspectratio=true, width=0.80\textwidth]{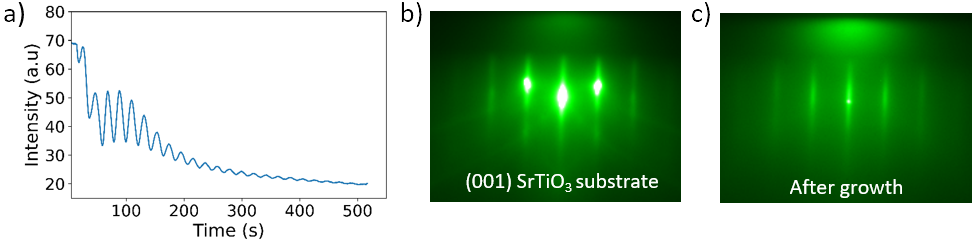}
		\caption{a) RHEED intensity oscillations observed during the PLD growth of Pr\textsubscript{0.8}Sr\textsubscript{0.2}NiO\textsubscript{3} perovskite thin films on SrTiO$_3{}$ (001) substrate. b) RHEED diffraction patterns observed before (STO substrate) and c) after deposition of the perovskite thin film.}
		\label{fig:S1}
	\end{figure*}

	\begin{figure*}[ht]
		\includegraphics[keepaspectratio=true, width=0.75\textwidth]{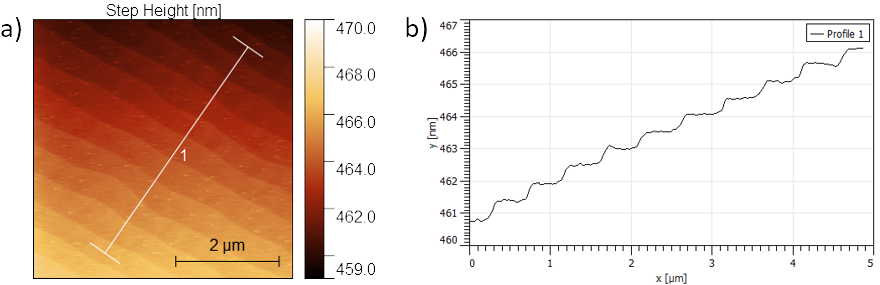}
		\caption{a) AFM image of the as-grown parent Pr\textsubscript{0.8}Sr\textsubscript{0.2}NiO\textsubscript{3} thin film on SrTiO$_3{}$ (001) substrate performed by atomic force microscopy. The white line indicates where the profile (shown in b)) was taken to estimate the width and height of the surface steps.}
		\label{fig:S2}
	\end{figure*}

	\begin{figure*}[ht]
		\includegraphics[keepaspectratio=true, width=0.75
		\textwidth]{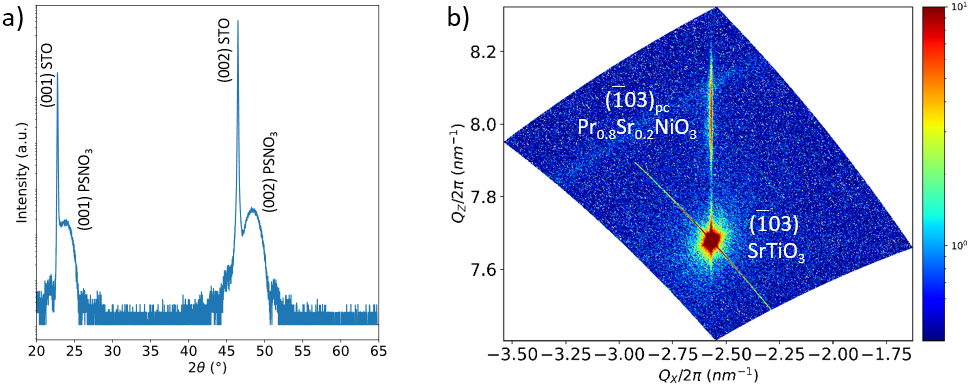}
		\caption{a) X-ray diffraction $\theta$-2$\theta$ scan of an as-grown perovskite PSNO$_3{}$ thin film, showing robust and intense (00l) pseudocubic perovskite peaks and no additional peaks.  b) High-resolution Reciprocal space map (RSM) around the ($\bar{1}$03) asymmetric reflection of a PSNO$_3{}$ film and STO substrate, demonstrating the epitaxial growth of the perovskite PSNO$_3{}$ thin films, with an in-plane lattice constant of 3.91 $\AA$.}
		\label{fig:S3}
	\end{figure*}

	\begin{figure*}[ht]
		\includegraphics[keepaspectratio=true, width=0.80\textwidth]{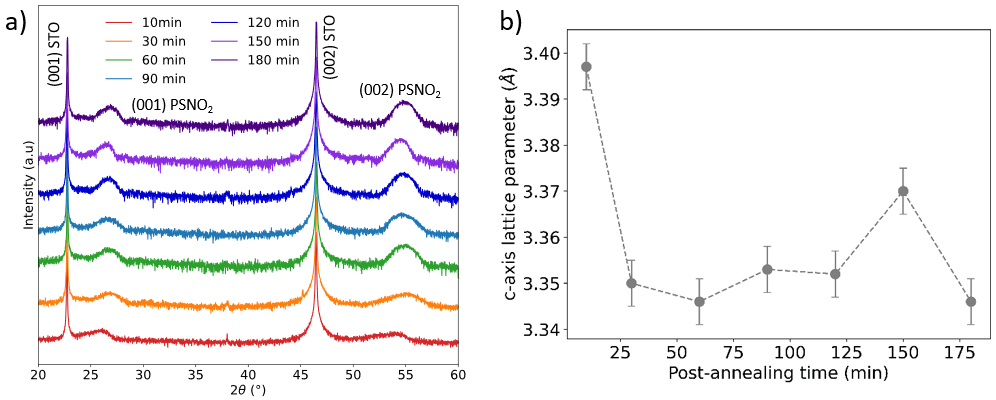}
		\caption{a) Comparison of XRD $\theta$-2$\theta$ scans for 8 nm thickness PSNO$_{3}$ sample reduced by Al deposition using different post-annealing time, ranging from 10 to 180 minutes. All the reductions were performed at fixed temperature of 380°C. b) c-axis lattice parameters ($\AA$) as a function of the post-annealing time from 10 to 180 minutes.}
		\label{fig:S4}
	\end{figure*}

	\begin{figure*}[ht]
		\includegraphics[keepaspectratio=true, width=0.85\textwidth]{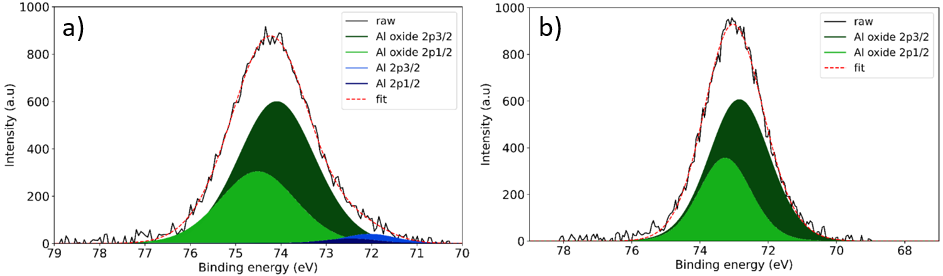}
		\caption{a) In-situ X-ray photoemission spectroscopy (XPS) of the reduced Pr\textsubscript{0.8}Sr\textsubscript{0.2}NiO\textsubscript{2} sample on SrTiO$_{3}$ (001) substrate after aluminum metal deposition (thickness = 2.8nm). The green peaks represent the aluminum oxide 2p3/2 and 2p1/2 peaks, respectively. The blue peaks represent the aluminum metal 2p3/2 and 2p1/2 peaks, respectively. The separation of the spin orbit splitting is fixed to 0.42 eV ± 0.1 eV. The peaks of aluminum oxide were fitted with a gaussian-Lorentzian function GL(30), whereas the peaks of aluminum metal were fitted with Lorentzian asymmetric function LA(1.53;243). Shirley method was used for background subtraction. Peak area ratios between components spin splitting were constrained to the theoretical ratio of 1:2. b) XPS of the same sample after exposition to the air.}
	\label{fig:S5}
	\end{figure*}

	\begin{figure*}[ht]
		\includegraphics[keepaspectratio=true, width=0.80\textwidth]{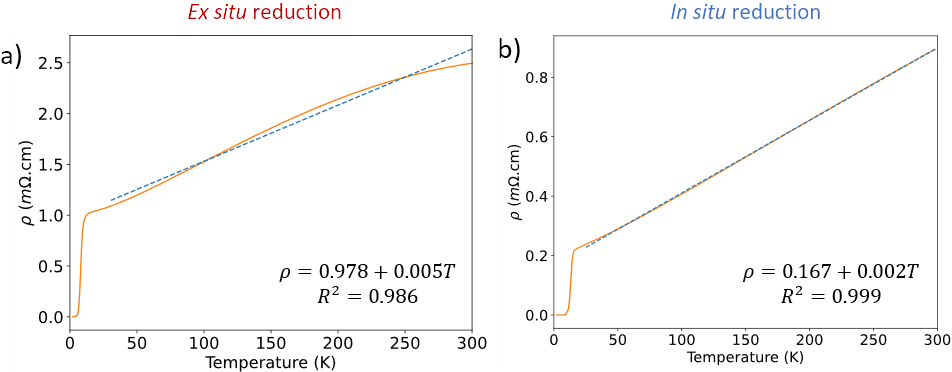}
		\caption{Linear fitting of the resistivity as a function of the temperature ($\rho (T)=\rho_{res}+AT)$) for ex-situ (a) and in-situ (b) reduced samples in te rage from 300K to 25 K.}
		\label{fig:S6}
	\end{figure*}

	\begin{figure*}[ht]
		\includegraphics[keepaspectratio=true, width=0.65\textwidth]{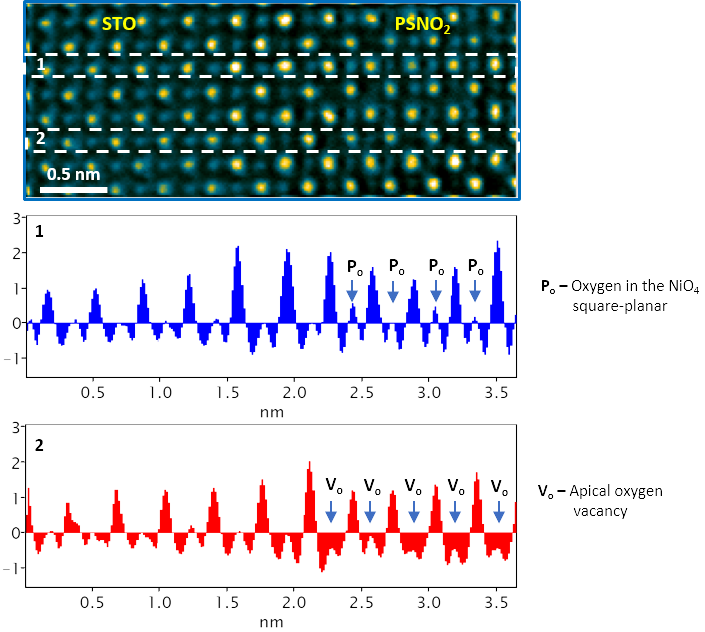}
		\caption{4D-STEM dCOM indicating the absence of apical oxygen atoms in the nickelate thin-film, near the interface PSNO$O_{2}$/STO substrate. The thickness of the film is around 6 nm. The profile 1 shows the oxygen in the Ni$O_{4}$ square-planar ($P_{O}$). The profile 2 shows the apical oxygen vacancy ($V_{O}$) in the infinite-layer phase.}
		\label{fig:S7}
	\end{figure*}

	\begin{figure*}[ht]
		\includegraphics[keepaspectratio=true, width=0.80\textwidth]{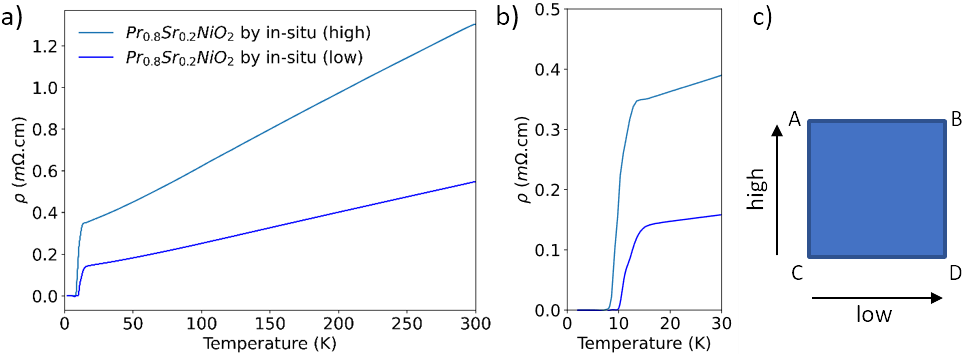}
		\caption{a) Resistivity measurement for an in-situ reduced thin film of 8 nm (5x5 size sample) in two different directions denoted as low (applied current A-B, voltage reading C-D) and high (applied current A-C, voltage reading B-D), showing variations in the resistivity amplitude and the transition temperature $T_{C}$. b) Expanded view of plot showed in panel a) near the transition temperature $T_{C}$. c) Sketch showing the different measurement directions.}
		\label{fig:S8}
	\end{figure*}

	\begin{figure*}[ht]
		\includegraphics[keepaspectratio=true, width=0.90\textwidth]{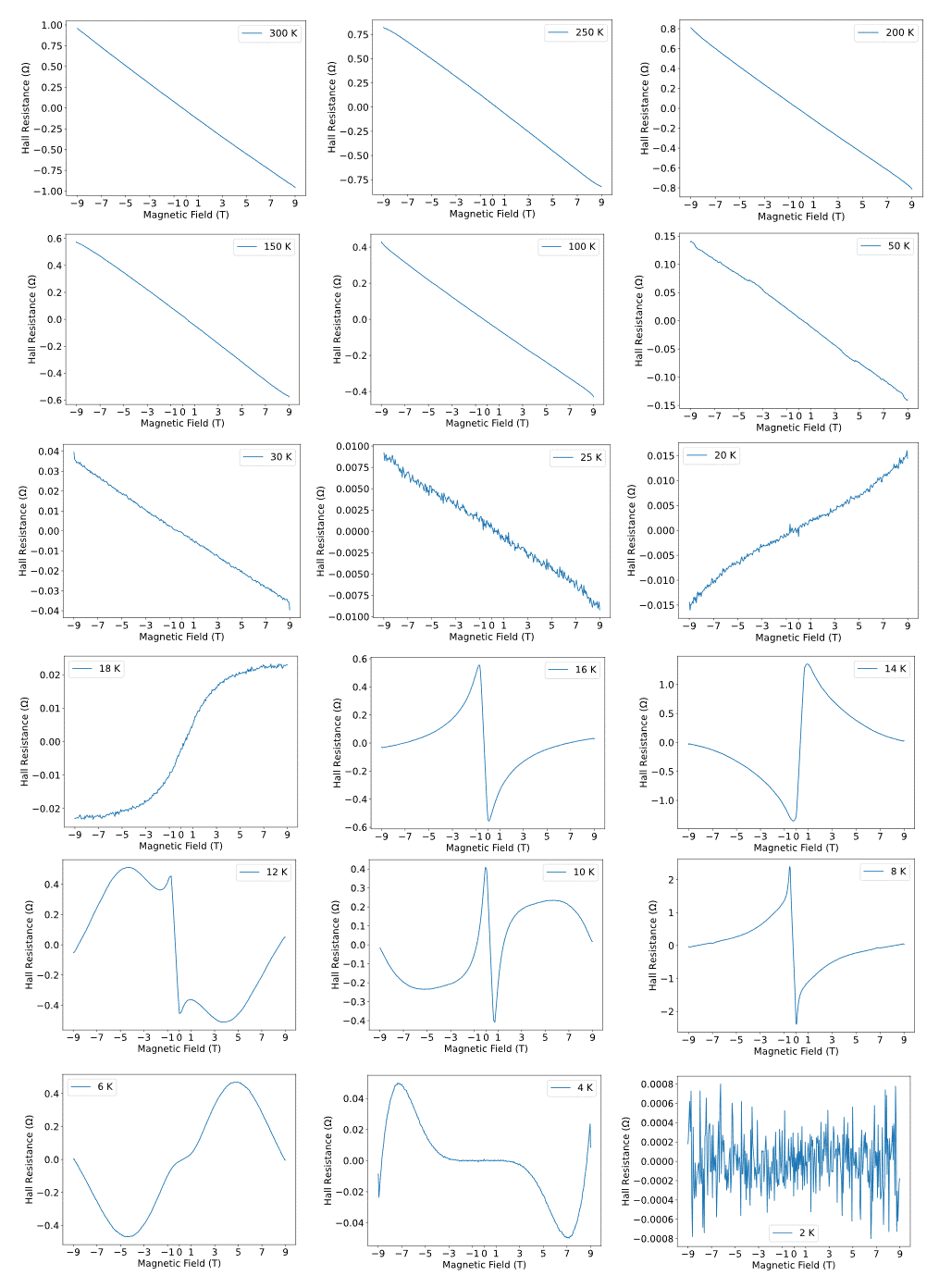}
		\caption{Anti-symmetrized Hall resistance as a function of the magnetic field at different temperatures for an in-situ reduced thin film of 8 nm thickness. Current was set to 0.1 mA at different temperatures.}
		\label{fig:S9}
	\end{figure*}

\end{document}